# EMPIRICAL STANDARDS
## for Software Engineering Research

Version 0.2.0

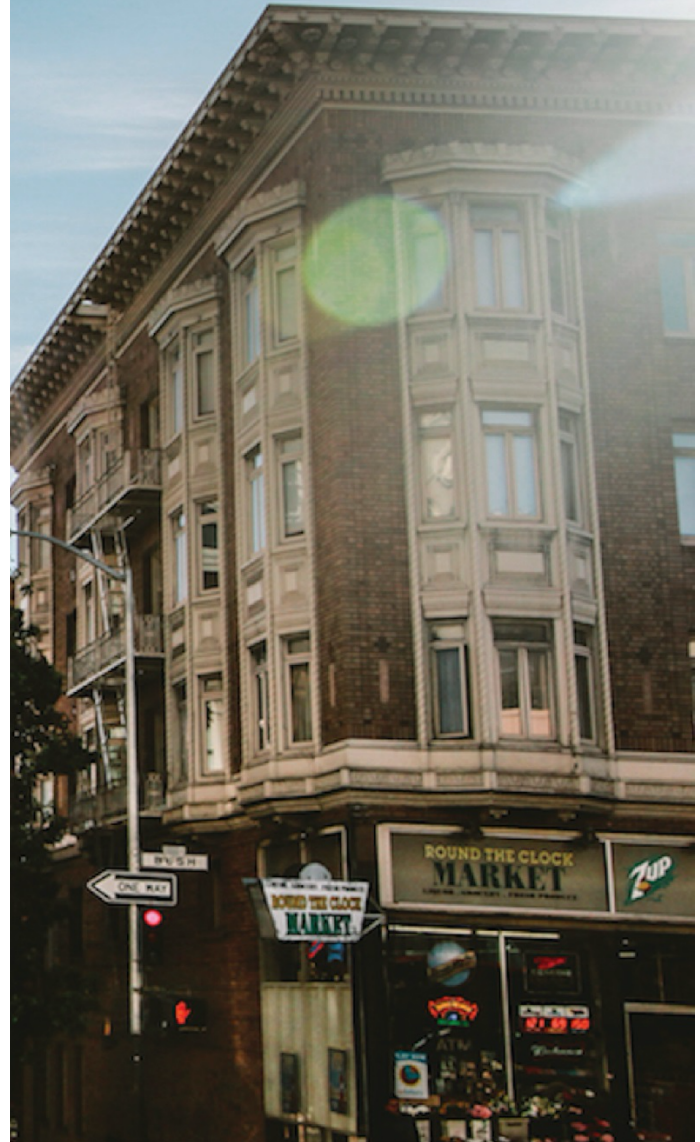

## MARCH 4

Paper and Peer Review Quality Task Force
Edited by: Paul Ralph

https://github.com/acmsigsoft/EmpiricalStandards

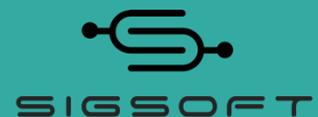

# Table of Contents

# Introduction

Scholarly peer review is crucial to science: it determines not only what is published where but also who is hired, funded and promoted. Yet, virtually no one is happy with the current state of peer review. Every academic has peer review horror stories ranging from frustrating to emotionally devastating. Empirical research consistently shows that peer review is unreliable and ineffective.[1] While there is no silver bullet to solve every issue with peer review, this report presents a plan for mitigating some of peer review's most serious problems.

Most journals and conferences provide general guidelines to authors and reviewers. Often, these guidelines ask reviewers to evaluate "soundness" but do not explain specifically what soundness means for an experiment, case study, survey, literature review, etc. Each reviewer must therefore construct method-appropriate evaluation criteria. Reviewers may not have the necessary background to construct appropriate criteria. Even if the reviewers are familiar with the method, constructing a new evaluation system for each paper is just too much work. Instead, reviewers tend to simply highlight problems that they notice. Even if a reviewer *does* create a more systematic scoring system, their system might be incompatible with that of the other reviewers, the editors, the authors, established methodological guidelines and community norms. From the venue's perspective, then, it is impossible to communicate expectations transparently. From the author's perspective, *the real evaluation criteria are secret.*

This disconnect between authors' and reviewers' expectations is why peer review is so unpredictable and frustrating. Meanwhile, the common disconnect between reviewer expectations, community norms and published methodological guidance makes peer review fundamentally unscientific. The truth is that constant rejection is neither intrinsic to science nor necessary for quality control. Rather, constant rejection is rooted in dissensus within scientific communities regarding how research should be conducted, and the accidental obfuscation of review criteria. Consequently, we can simultaneously increase review quality, paper quality and acceptance rates by generating and evolving *empirical standards*.

# Empirical Standards

> ***Empirical Standard:*** *A brief public document that communicates expectations for a specific kind of study (e.g. a questionnaire survey).*

The best way to understand what we mean by an empirical standard is to look at a standard for a familiar methodology. The standards are hosted on the Empirical Standards GitHub repo.[2] (You can also report issues and suggest changes via the repo—more on that later).

Empirical standards **are not** vague criteria like "soundness" and "presentation." Each standard lists the specific attributes we expect for a particular methodology, e.g., "uses random assignment" (experiment) or "presents clear chain of evidence from interviewee quotations to proposed concepts" (qualitative survey).

---

[1] For summary, see:
Ann Weller, 2001. Editorial peer review: Its strengths and weaknesses. *Information Today, Inc.*
P. Ralph (2016) Practical suggestions for improving scholarly peer review quality and reducing cycle times. *Communications of the Association for Information Systems,* (38), Article 13.
[2] https://github.com/acmsigsoft/EmpiricalStandards

Empirical standards **do not** replace expert judgment with inflexible rubrics. For example, if researchers do not report effect sizes with confidence intervals because they took a Bayesian approach and report posterior probabilities instead, they just say so. Each standard gives examples of "acceptable deviations" to reinforce the fact that research can deviate from the standards for good reasons. Furthermore, empirical standards focus on the methodological substance of a study; they do not micromanage style.

An empirical standard is essentially a one-page checklist of specific criteria that can be used by authors to conduct and report research, and by reviewers to evaluate manuscripts.

Adopting empirical standards makes peer review *fair* and *transparent* in two ways. First, the standards allow the authors and the reviewers to work from the same rubric, so there are no secrets and fewer surprises. Second, the standards allow crucial decisions about what is and is not acceptable scientific practice to be made by our scientific community *collectively*, instead of by reviewers, *individually*.

## Components of the Standards

There are three kinds of standards. **The General Standard** applies to all empirical research. We wrote a general standard to reduce duplication and fall back on when no more specific guidance is available. For example, **The Case Study Standard** does not say "states a clear research question" because that is already in **The General Standard**.

Next, there is a set of methodology-specific standards including **The Experiment (with Human Participants) Standard**, **The Questionnaire Survey Standard**, and **The Case Study Standard**. Each methodology-specific standard includes the following items.

- Name of method (e.g. Experiment, Questionnaire Survey, Case Study)
- Definition of the method
- Application (i.e. where the standard does and does not apply)
- Specific Attributes (i.e. the properties of a manuscript that make it publishable), organized into three categories:
    - *Essential:* necessary conditions for publishing the work. Without a compelling justification, a study that does not meet one or more essential attributes should not be published in any scholarly venue.
    - *Desirable:* attributes that are recommended but not always necessary or applicable. Some desirable attributes may be mutually exclusive. For publication in a prestigious venue, a study should exhibit some, *but not all*, desirable attributes.
    - *Extraordinary:* attributes that are not required even for the most prestigious venues, and typically indicate award-quality research
- General Quality Criteria (e.g. internal validity, construct validity, credibility, resonance)
- Examples of Acceptable Deviations (to emphasize that standards are flexible)
- Antipatterns (i.e. common problems with this methodology)
- Invalid criticisms (i.e. unreasonable arguments against a paper that reviewers should not make)
- Suggested readings (including longer methodological guidance upon which standards are based)
- Exemplars (good examples of the method authors should emulate; useful for teaching grad students)
- Notes (important clarifications that did not fit elsewhere)

Finally, we have also included several **Supplements** for cross-cutting concerns like information visualization and sampling. These supplements help to reduce duplication between the standards and give more information about complex issues.

## Interpreting and using the Standards

The standards can be used in several ways:
- Scholarly venues can adopt the standards to communicate their expectations more precisely to both authors and reviewers (see **Adoption Models**).
- Reviewers can use the standards to evaluate manuscripts.
- Researchers can consult the standards when designing their studies.
- Authors can use the standards as a pre-submission checklist to ensure that their manuscripts explicitly address methodological issues reviewers are likely to examine.
- Educators can use the standards to inform the curricula for their research methods courses. Again, the standards point to rather than replace the research methods discourse, but they are simultaneously broad and concise, and the lists of recommended readings and exemplars are an excellent resource for delivering advanced research methods courses.

From a reviewing perspective, the empirical standards are intended for evaluating *complete, empirical studies*. The standards are not intended to review pilot studies and non-empirical scholarship. (Venues should consider scaling back peer review for pilot studies, posters, short papers and workshop papers and position papers to control reviewing loads.)

The standards are modular to reduce duplication, so most manuscripts should be evaluated using multiple standards. For instance, a controlled experiment with several charts and graphs might be evaluated using **The General Standard**, **The Experiment Standard**, and **The Information Visualization Supplement**. A multimethodological study combining a systematic literature review with a questionnaire survey would be evaluated against **The General Standard**, **The Systematic Review Standard**, **The Questionnaire Survey Standard** and **The Multi-Methodology Supplement**. A system will automatically assemble the relevant standards into a review form rather than each reviewer trying to manually select and combine appropriate standards.

If a manuscript uses a method for which no specific standard exits, reviewers should jot down what they believe are the essential attributes for this method based on their knowledge of published methodological guidelines and community norms. They should then evaluate the manuscript using **The General Standard**, which applied to all empirical research, and their own list of essential criteria. Obviously, this approach is not ideal, but it's the best we can do until an appropriate specific standard can be generated.

## Two Crucial Rules for Reviewers

Reviewers using the empirical standards should obey two critical rules:

1. **BE FLEXIBLE.** Do not play "gotcha" with the standards, looking for any excuse to reject a paper. Whenever a criterion is not met, ask yourself two questions: (a) "Does deviating from the standards in this way make sense in the context of this study?" and (b) "Would this problem be easy to fix in the camera-ready copy?" Justified deviations and easy-to-fix mistakes should not preclude acceptance.

2. **NO NEW CRITERIA.** The absence of an attribute indicates that it should not be considered. For example, **The General Standard** does not say that papers should be free from arbitrary decisions because most research involves some arbitrary decisions and their presence does not invalidate findings. Do not apply your own, non-standard criteria to a manuscript. If you find a critical flaw that isn't captured by the standards, please lodge an issue on the empirical standards repo (without unmasking yourself as a reviewer).

# Design Process

In 2018, ACM SIGSOFT solicited volunteers, using the SEWORLD mailing list and social media, for a series of special initiatives. Inclusion was not limited to SIGSOFT members. Paul Ralph (*Dalhousie University*) and Romain Robbes (*Free University of Bozen-Bolzano*) were asked to co-chair the *Paper and Peer Review Quality Initative*. Paul used the initiative to explore the idea of empirical standards.

After the initiative leaders were announced at the ICSE town hall in May 2019, Paul contacted the volunteers and shared plans for drafting empirical standards. Over 50 people eventually joined the *Initiative.* Volunteers:

- reviewed the plan for drafting the standards;
- discussed which methods should have a standard;
- created a template illustrating the sections each standard should have;
- suggested additional researchers who might want to volunteer; and
- nominated subject matter experts to draft each standard.

Each standard was assigned to a team of subject matter experts who produced a draft, based on existing published guidance where possible. The drafts were then edited for consistency, condensed, compiled and circulated (internally within the initiative) for feedback, and updated accordingly. Next, we created a GitHub repository, posted the first draft of this report and the first eight standards, and solicited feedback from the SE research community. Meanwhile, we recruited more subject matter experts to create the rest of the standards, and transitioned all completed standards to plain text (GitHub markdown) to facilitate version control and pull requests.

Consequently, the empirical standards have been generated by a large, diverse team of domain experts (see **List of Contributors**). Current versions of the standards are available on the GitHub repo.

## Design Principles
The standards are guided by the following principles.
- Manuscripts should be evaluated on their own terms, vis-à-vis their context and goals.
- The scientific community (not individual reviewers) should determine expectations and norms.
- Expectations of journals and conferences should *reflect* community consensus.
- Researchers should follow standards *where they make sense and justify reasonable deviations*.
- Reviewers should focus on the manuscript's methodological soundness.
- Reviewers should **not** micromanage style or copyedit manuscripts.
- Reviewers should focus on a study's use of best practices (e.g. using validated scales), **not** abstract criteria (e.g. construct validity)
- The standards *must not be biased* against specific research methodologies or areas of interest.

## Evolution and Governance

The empirical standards are not final. They are living documents, which should be continuously refined to better reflect the views of our community and revised to evolve with those views. Issues can be reported and changes can be requested via the GitHub rep. It is crucial that the standards are fair and inclusive. Any active software engineering researcher can suggest improvements. We envision that, as the standards are used, authors and reviewers will be highly motivated to report problems and suggest improvements, and so the standards will evolve rapidly.

Once completed, each standard will require one or more *maintainers*. Each maintainer will be responsible for reviewing issues and change requests filed against their standard, and will have the authority to make, accept or reject changes. Crucially, maintainers must be experts *in that kind of research* to avoid misapplying norms from one kind of study to a fundamentally different kind. To become a maintainer, therefore, a person must:

- Have a PhD or equivalent terminal degree in software engineering, computer science or a related discipline.
- Have published one or more studies related to the standard (e.g. for the systematic reviews standard, a maintainer must have published a paper that reports a systematic review, analyzes a sample of systematic reviews, or provides guidelines for performing systematic reviews) in a prestigious software engineering journal or conference in the past six years.
- Be approved by the *steering committee*.

The *steering committee* will consist of:

- The director of the empirical standards project—currently Paul Ralph (Dalhousie University)
- The editors-in-chief (or their designated representatives) of all of the DBLP-indexed journals that adopt the standards
- A designated representative of the steering committee of each DBLP-indexed conference that adopts the standards
- Two, elected, early-career members-at-large

The steering committee will determine how it appoints maintainers and can institute policies to exclude illegitimate or predatory venues or include someone who, in hindsight, should have been included. Once the initial steering committee is formed, it will be responsible for elaborating and formalizing the governance model.

# Adoption Models

Academic venues can begin using the standards in at least three different ways: accidentally, partially or fully. The more fully venues adopt the standards, the faster they will drive improvements in both peer review and paper quality.

## Accidental Adoption

Many reviewers will begin using the empirical standards on their own, whether or not a venue instructs them to do so, because the standards make reviewing easier. The standards reduce the extraneous cognitive load associated with generating ad hoc criteria for each review and tell reviewers what to look for in unfamiliar methods. Meanwhile, researchers will increasingly use the standards to design better studies and write better manuscripts because the standards can be read and digested much faster than whole books and papers, and because reviewers are starting to use them. Therefore, the standards will begin (and are already beginning) to infiltrate review processes even if venues ignore them.

Accidental adoption may cause conflict. Imagine having a manuscript rejected for failing to comply with standards you have never heard of. You might feel mistreated by the venue. You might reasonable ask "If you were going to evaluate my manuscript based on a standard, why didn't you tell me up front?" Similarly, reviewers referencing the standards might be admonish by other reviews (or editors, etc.) who don't know of them or don't like them. Moreover, if reviewers have questions regarding the interpretation or application of the standards, they will not know who to ask or how to get support.

Venues can avoid these conflicts by formally adopting the standards.

## Partial Adoption

Venues can tentatively adopt the standards by explicitly encouraging authors and reviewers to use them. We recommend the following steps, and can help with their implementation:

1. Add references to the standards to the author guidelines and reviewer directions.

2. Add a question to the submission system such that authors can indicate the research method(s) used in their papers by selecting from a list based on the standards. Make sure reviewers will receive this information.

3. Offer training to the reviewers on how to interpret and apply the standards consistently.

4. Change the wording of the recommended decision field of the review form as shown in Table 1 (journals) and Table 2 (conferences). Note that we recommend offering just three choices and avoiding asking directly for accept or reject recommendations. Rather, whether to accept or reject is a consequence of the manuscript's relationship to the standards.

| **Old** | **New** |
|---|---|
| Reject | The paper does not have one or more essential attributes AND<br>One or more deviations are not justified AND<br>One or more unjustified deviations cannot be fixed, or at least not without repeating data collection |
| Invite Revision | The paper does not have one or more essential attributes AND<br>One or more deviations are not justified AND<br>All of the unjustified deviations can be fixed without repeating data collection |
| Accept | The paper has all the essential attributes OR<br>All deviations are justified or trivially fixed |

*Table 1: Recommended Decision Categories for Multi-stage Review Processes (most Journals)*

| Old | New |
|---|---|
| Reject | The paper does not have one or more essential attributes; AND One or more deviations are not justified AND One or more deviations cannot be fixed by modest editing alone[3] |
| Accept | The paper has all the essential attributes OR All deviations are justified or can be fixed by modest editing alone[3] |
| Accept & Nominate for distinguished paper | The paper has all the essential attributes (or deviations are justified) AND the paper has several desirable or extraordinary attributes. |

*Table 2: Recommended Decision Categories for Single-stage Review Processes (most Conferences)*

The main benefit of this approach is that it is easy to implement without major changes to editorial management services like EasyChair and HotCRP. The main problem with this approach is that some reviewers will continue going rogue—disregarding community-generated expectations in favor of their own made-up, unvalidated criteria. If authors meticulously craft studies and manuscripts to comply with standards that reviewers ignore, the review process will seem even more unjust and enraging. Therefore, venues should consider a more structured review process.

## Full Adoption (Structured Review)

Currently, we review papers like critics review films. We watch/read the film/paper and write an essay expressing our views of its quality. But we are scientists evaluating studies, not critics evaluating films. Writing a rambling essay is not the best way to evaluate a scientific study. Venues can replace the essay with a more structured review form, generated from the standards, with more limited free-text fields. The general process is as follows.

1) The venue sets decision rules (like those shown in Table 1 and Table 2).
2) Authors indicate which research method(s) their submission uses.
3) The empirical standards system generates an *author's pre-submission checklist*. The checklist contains the same attributes the reviewers will use to evaluate the paper.
4) Authors can use the checklist to make sure that their manuscript clearly shows how it complies with the standards and justifies deviations.
5) The authors submit their manuscript.
6) Someone appointed by the venue (e.g. managing editor, submissions chair) completes the initial checks listed in **The General Standard** and verifies that the correct standards have been selected. If not, they seek clarification from the authors and correct the method metadata.
7) The empirical standards system generates a dynamic review form (for reviewers)—see below.
8) Two reviewers read the manuscript and answer the questions.
9) *(optional)* Reviewers resolve specific disagreements (e.g. whether the paper checks assumptions of statistical tests) by discussion. Reviewers do not discuss whether to accept or reject the paper because they are not making accept/reject recommendations directly. The system makes the recommendation based on the reviewers' answers.
10) *(optional)* A third reviewer is recruited if the first two reviewers cannot reach a specified agreement threshold (does not have to be 100%; does not apply to all questions).
11) Reviewers do not make an overall recommendation (e.g. accept, major revision). Rather, the system compiles the reviewers' answers and tells the editor whether the paper meets the venue's rules for acceptance, revision or rejection.

---

[3] It would be wise to give some examples. For instance, "modest edits" might include rewriting the limitations section of the paper, swapping out a Pearson correlation for a Spearman correlation, or adding task materials to a replication package. Collecting new data, adding entirely new analyses, or overhauling the structure of a paper would not be "modest edits."

12) Based on the standards, the reviewers' answers and the venue's decision rules, the system generates and sends to the authors:
    a) a review summary (if the paper is accepted)
    b) a to-do list for the authors (if a revision is invited)
    c) a summary of the reasons for rejection (if the paper is rejected)
13) *(optional)* In addition to accept-reject decisions, the system could be configured to assign various badges or grades to accepted manuscripts (e.g. distinguished paper award, artifact-available badge).

The standards provide specific attributes as static checklists. The problem with static checklists is that they might encourage unthinking rejection of good research that is just a little bit different. To correct this, we propose a dynamic review form that looks like a checklist until the reviewer indicates that an essential attribute is missing, and then reveals diagnostic follow-ups. The system can use these follow-ups to determine the accept-reject decision and the content of the decision letter (e.g. a list of necessary revisions or reasons for rejection), as shown in Figure 1 and Figure 2.

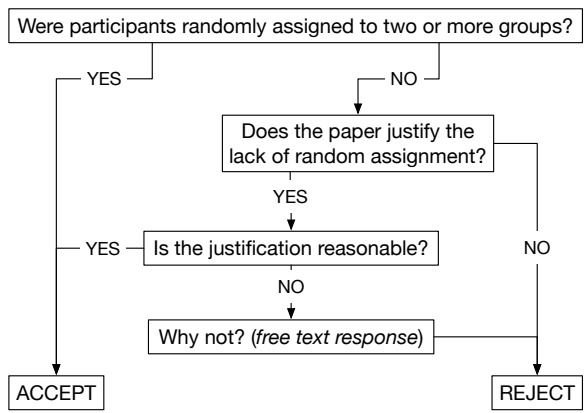
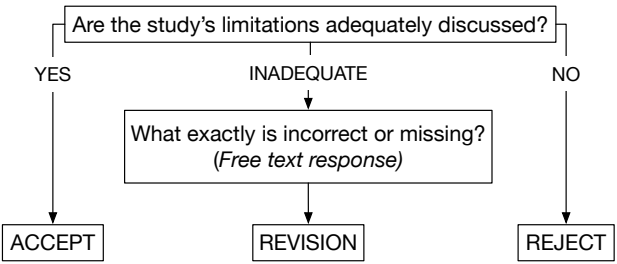

*Figure 1: Random Assignment Decision Tree*    *Figure 2: Limitations Decision Tree*

Figure 1 illustrates a series of questions for random assignment (from the **Experiments with Human Participants Standard**). By specifically asking reviewers whether there is a reasonable justification for lack of random assignment, standards-based review can balance the competing needs for structure and flexibility.

Decision trees can also help manage problems that are important but easily fixed in revisions. For example, Figure 2 shows possible pathways for evaluating a paper's limitations section. If the paper does not even attempt to explain the study's limitations, we reject. If the limitations section is adequate, we accept. If, however, the limitations section is inadequate, we ask the reviewers exactly what is incorrect or missing and invite a revision. This format encourages the reviewer to give actionable advice and transfers to the venue decisions regarding what problems lead to revision or rejection.

Again, a decision tree is only shown when reviewer indicates that an essential attribute is missing. When a paper complies with the standards, the review will simply tick all the 'yes' boxes; follow-up questions only appear when a 'no' box is ticked.

In general, the reviewer only enters text to justify a criticism or describe something needed for a revision. Such text should be brief. Space for free-form comments can be included at the end of the form, but such comments should be optional and much shorter than the review-essays common today. Critically, free-form comments do not affect the accept-reject decision, the revision to-do list, or the evaluation of any revisions.

Regarding revisions, full adoption of empirical standards should increase acceptance rates and decrease the frequency and scope of revisions. The fact that virtually every journal paper requires a major revision is not caused by the intrinsic difficulty of writing good papers but by the opacity and inconsistency of current review processes. Both major revisions and rejections should be rare. Empirical standards will reduce the need for revisions and rejection by addressing the main reason journal papers are never accepted on the first round: the disconnect between the mental models of the authors and reviewers.

When revisions are needed, the authors will get a clear to-do list generated by the system, perhaps with a little finessing by an associate editor, discussant, etc. The revision to-do list is entirely based on missing essential attributes. There is no mechanism for mandating other changes. This prevents reviewers from imposing ad hoc criteria. Because the changes are more clear-cut, a single person (e.g. associated editor, lead reviewer) can simply check whether the authors completed to to-do list. There is no need for multiple rounds of revisions or re-review.

To summarize, we envision a structured review process where reviewers evaluate a manuscript against a standard by answering predominately yes-or-no questions. A manuscript's accept-reject decision is a consequences of the decision rules devised by the venue (or default rules devised by the Empirical Standards Initiative). The system actively cues reviewers to evaluate the reasonableness of deviations and actively prevents researchers from imposing their own, non-standard criteria. The system reduces review loads by reducing revisions and obviates re-review.

# Cost and Benefits of Empirical Standards

An empirical standard is a model of a community's shared understanding of how a kind of study should be done. Providing these models to authors, reviewers and editors while implementing standards-based review should produce many benefits.

Up until now, the criteria used by academic venues have been so vague that every reviewer effectively has to invent their own rubric. From the author's perspective, therefore, *the real evaluation criteria are secret.* Creating the empirical standards allows academic venues to communicate their evaluation criteria transparently to both authors and reviewers.

## Benefits for Editors / Program Chairs

- improved review quality, thoroughness, consistency and readability
- reviewers cannot sit on the fence – the system always produces a clear recommendation
- more control over the review process and paper expectations; fewer rogue reviewers
- fewer angry emails from rejected authors
- easier to analyze inter-reviewer reliability and identify problem reviewers
- higher acceptance rates; faster publication times
- fewer low-quality submissions
- venues are perceived as being more fair, impartial and unbiased
- easier to recruit reviewers because reviews are less work

## Benefits for Reviewers

- less overall reviewing work
    - reviewers do not have to construct expectations, write an essay or copyedit the paper

- - less discussion among reviewers is needed; discussions can be more focused on disagreements between reviewers
  - third reviewer only needed if first two reviewers have insufficient agreement
  - fewer obviously deficient submissions because expectations are known
  - less re-reviewing manuscripts because more manuscripts are accepted straight away; revision to-dos are clearer, and most revisions can be checked by a single editor
  - all work that can be done by one person (e.g. checking keywords) is, and that one person is not a reviewer
  - less bouncing around of manuscripts between venues because many venues use the same standards and rejection rates are lower
- venue's expectations, for both reviewers and submissions, are clearer
- reviewers need less methodological expertise to review a paper effectively
- less arguing among reviewers: reviewers only have to determine whether a paper complies with a standard, not how, in principle, the study should have been conducted

## Benefits for Researchers

- higher acceptance rates because expectations are known in advance
- fewer major revisions because expectations are known in advance
- faster publication times because there are fewer major revisions
- reviews are less emotionally charged, upsetting and demoralizing
- reviews are more actionable and consistent; no mixed messages
- reviewers cannot micromanage style
- reviewers cannot impose requirements inconsistent with established norms and guidelines
- venue's expectations are transparent to authors; reviews are more predictable
- no more superficial reviews that do not engage with the paper's methodology and core claims
- standards provide convenient checklists for designing studies and polishing manuscripts
- standards mitigate bias against qualitative, exploratory and industry-focused research
- stops cross-paradigmatic criticism (e.g. "why doesn't this qualitative study have numbers?")
- standards are useful for training graduate students
- standards can be used in systematic literature reviews to assess the quality of primary studies

## Benefits to Society

- better, more rigorous, more ambitious studies
- less money and resources wasted because researchers forgot or ignored critical practices
- less HARKing and publication bias, because statistical significance is not a criterion
- less resistance to open reviewing / signed reviews (because reviews are less personal)
- less money and resources wasted on bad reviewing and re-reviewing
- fewer scientists and graduate students burning out due to destructive reviewing

## Eliminating Frustration

Peer review is so frustrating to authors not because reviewers find mistakes but because reviewers apply criteria that authors did not anticipate or do not agree with. Similarly, peer review can be frustrating for reviewers when authors omit or glaze over essential research practices. Likewise, peer review is frustrating for editors when reviews are superficial or wrong: the authors expect the editor to do quality control, but the editor worries that over-ruling reviewers looks biased. Empirical standards address all these frustrations. Reviewers cannot make up unexpected criteria, authors cannot sidestep essential criteria and editors will always get detailed, technical reviews because the structured review process simply doesn't allow any other alternative.

> *Myth: peer review is upsetting because reviewers find authors' mistakes*
> *Truth: peer review is upsetting because authors and reviewers disagree on appropriate research practices*

Meanwhile, from a methodological and philosophical viewpoint, peer review is frustrating because (1) reviews often focus on style over substance; and (2) broad debates about how research should be conducted play out within the secret review processes of individual manuscripts instead of within the community at large. Structured review with empirical standards addresses the first frustration by focusing reviewers on substance. Meanwhile, extracting the process of creating the standard from the process of reviewing a specific paper facilitates a broad debate about how studies should be performed and presented. Reviewers implement the collective views of the community, rather than imposing their individual opinions.

## Costs

The main cost of this project is building and maintaining the standards and the system that generates the review forms. Luckily, many scholars care deeply about improving peer review, and dozens of volunteers have been generous with the time and expertise to draft the standards. A [prototype structured-review system](https://web.cs.dal.ca/SIGSOFT-Empirical-Standards/)[4] is under development, but more help will be needed here.

As explained above, partial adoption of the standards requires only superficial changes to the review process. Full adoption requires more fundamental changes, and we will need to collaborate with review process management systems including Easychair and HotCRP to implement these changes.

## Challenges with Structured Review

Empirical standards *direct*—rather than *replace*—expert judgment. The standards use words like "reasonable" and "convincing." A human being still has to judge whether things are reasonable and convincing. Using Empirical Standards neither makes reviewers into robots nor enables automating the review process. Empirical standards direct reviewers in the same way that laws direct judges.

Empirical standards, if implemented poorly, could lead to a superficial, inflexible, box-ticking approach to review. As explained above, the standards include examples of acceptable deviations and the review forms will include follow-up questions like "does the paper justify this deviation" and "is this deviation reasonable in the context?" to balance the competing needs for structure and flexibility.

Another possible problem is that reviewers may believe that a paper should be accepted despite violating the standards. This problem too is solved by our follow-up questions. Reasonable deviations from the standards are acceptable, so the reviewer simply indicates that the deviation is reasonable.

The opposite problem is trickier. What if reviewers believe the paper should be rejected despite complying with the standards? In other words, the reviewers think there is a problem with the standards. Here, the desired workflow is to report the problem in the standards issue tracker (see below) so the standard can be updated, and, meanwhile, *accept the paper*. The paper should be accepted for three reasons:

1. Rejecting papers over problems not mentioned in the standards is unfair to the authors

2. If some critical flaw in one of the standards allows heaps of invalid research through, it will be quickly found and corrected.

3. Allowing reviewers to override the standards at will would be rife for abuse. Reviewers would carry on inventing bogus, non-consensus criteria. If the average reviewer could be so trusted, peer review would not be in its current, dismal state.

---

[4] https://web.cs.dal.ca/SIGSOFT-Empirical-Standards/

Finally, the standards could create unforeseen problems. To mitigate unknowns, we should conduct pilot studies at conferences and journals, and continually evaluate the ongoing impact of standards and structured review, where they are applied. We can and should empirically study how the standards affect the review process.

# Justice, Equality, Diversity and Inclusion

The current system of peer review is demonstrably biased in many ways.[5,6] One purpose of developing these empirical standards is to make peer review less biased. However, the more a person has suffered from these biases, the less likely they would be involved in creating and implementing these standards. To mitigate bias, we took the following steps:
- there was an open call for volunteers, published widely via the SEWorld mailing list, and discussed at multiple conferences and on social media
- anyone can report an issue or make a suggestion for improving the standards
- we specifically tried to recruit women, people of color and members of other underrepresented groups to be involved with the standards
- more than 40 subject matter experts were involved in drafting the standards
- different methods have different standards, reducing bias against specific approaches
- we tried to make the general standard, especially, philosophy- and method-agnostic
- we are using GitHub to make all feedback and changes transparent
- we added a provision for members at large to get a junior perspective on the steering committee

These steps are likely insufficient to eliminate bias in the standards but are the best we could come up with. Further suggestions are welcome at any time.

## The Scientific Basis of Empirical Standards

There exists a class of studies in which one or more properties of the unit of analysis are measured using expert judgement. For example, suppose we want to know whether developers write better software documentation when using a particular tool that generates (partial) documentation from source code. We randomly divide a sample of developers into two groups. The control group writes documentation from scratch; the treatment group uses the tool and modifies its output.

---

[5] Giangiacomo Bravo, Mike Farjam, Francisco Grimaldo Moreno, Aliaksandr Birukou, and Flaminio Squazzoni. 2018. Hidden connections: Network effects on editorial decisions in four computer science journals. *Journal of Informetrics* 12, 1: 101-112.
[6] Paul Ralph. 2016. Practical suggestions for improving scholarly peer review quality and reducing cycle times. *Communications of the Association for Information Systems* 38, 1: article 13.

To measure our dependent variable—documentation quality—we have an expert in software documentation rate each document on a five-point scale. Then we test the hypothesis that the quality of the treatment group is higher than the quality of the control group.

But how do we know that our quality ratings are *reliable*? We don't. So instead of one rater, we get two or three raters, and we estimate inter-rater agreement using a statistic like Cohen's Kappa or Krippendorff's Alpha.[7] When agreement among raters is high, we are more confident in the ratings.

We know that peer reviewers exhibit low inter-rater agreement.[8] How many journal submissions have you gotten back where one reviewer says "minor revision", another says "major revision" and the third says "reject"? That kind of disagreement is supposed to be *rare*. Now some people think "yeah but scholarly peer review is different." No, it isn't. That's the special pleading fallacy. If reviewers can't agree, peer review is capricious and unscientific.

Furthermore, scientists figured out how to improve inter-rater reliability decades ago.[9] You do not complain that the raters are stupid and give up. You keep defining *decision rules* to guide raters until sufficient agreement is reached. In other words, you increase task structure until decisions become consistent.

Peer review isn't unreliable because reviewers are ignorant or lazy or pedantic or whatever we yell at our screens when we don't like a review. Peer review is unreliable because it lacks structure. 'Write an essay and then state a recommendation' isn't structure. 'List strengths and weaknesses' isn't structure. A review form with 25 true-false questions about specific methodological practices is *structure*. The empirical standards will make peer review reliable by adding structure in the same way that a coding scheme and decision rules make expert judgment studies more reliable.

Meanwhile, a significant body of research demonstrates that checklists improve consistency and performance in diverse settings including research reporting[10] and healthcare[11].

Some people will resist structured review because they like the freedom of essays. Unfortunately, essay-review gives reviewers the freedom to reject good work for invalid reasons and to impose unreasonable demands upon authors. Finding a better power balance is necessary to keep our community healthy and productive. By giving up a little freedom in our role as reviewers, we empower ourselves in our roles as authors, editors and members of a scientific community.

# Conclusion

Empirical standards are brief, concise models of a scientific community's expectations for different kinds of studies. They facilitate a radical transformation of peer review from a free-form essay to a structured evaluation, which will produce a more effective, reliable, transparent and fair peer review process, leading to better research, higher acceptance rates and lighter reviewing workloads. In particular, empirical standards will mitigate the frustration caused be differences between authors' and reviewers' expectations and transfer the power to set expectations from individual reviewers to our community as a whole.

---

[7] See the **Inter-Rater Reliability and Agreement Supplement**.
[8] Eric Price 2014. The NIPS experiment. https://blog.mrtz.org/2014/12/15/the-nips-experiment.html
[9] R. J. Bullock and M.E. Tubbs, 1987. The case meta-analysis method for OD. *Research in organizational change and development*, 1, pp.171-228.
[10] e.g., Amy C. Plint, David Moher, Andra Morrison, Kenneth Schulz, Douglas G. Altman, Catherine Hill, and Isabelle Gaboury. "Does the CONSORT checklist improve the quality of reports of randomised controlled trials? A systematic review." *Medical journal of Australia* 185, no. 5 (2006): 263-267.
[11] E.g. Jonathan R. Treadwell, Scott Lucas, and Amy Y. Tsou. "Surgical checklists: a systematic review of impacts and implementation." *BMJ Quality & Safety* 23.4 (2014): 299-318.

Make no mistake, implementing empirical standards and structured review involves *radical*, not incremental, change. Virtually everyone is unhappy with the current state of peer review, and our entire community is clamouring for change. Peer review is fundamentally broken. It cannot be fixed by fiddling around the edges. Only by redesigning peer review as a process of applying community-driven standards can real progress be made.

You can help advance the empirical standards project by:
- encouraging your favourite venues to adopt the standards
- submitting to and supporting venues that adopt the standards
- reading the standards and suggesting improvements
- referring to the standards in your reviews and manuscripts
- using [pre-submission checklists](#) to improve your papers
- using the standards to write better peer reviews
- using the standards to assess study quality in systematic literature reviews

## Disclaimers

Nothing in these standards should be interpreted as privileging any methodology (e.g. controlled experiment vs. ethnomethodology), epistemological position (e.g. positivism vs. interpretivism), analysis approach (i.e. frequentist vs. Bayesian statistics), or data type (e.g. quantitative vs. qualitative). Each study should be reviewed on its own terms, which is why so many different standards are needed.

## Acknowledgements

The editor would like to acknowledge:
- the foresightedness of ACM SIGSOFT, and President Thomas Zimmerman in particular, for kickstarting this initiative
- the intellectual generosity of all the subject-matter experts who contributed content to the standards
- the competence and diligence of all the technical contributors who organized the repo and created the prototypes
- the moral courage of the editors and conference organizers who became early-adopters of the standards

# Glossary

**Anonymization:** the process of removing or transforming identifying elements of data such that it cannot be used by a third-party or the data-custodian themselves to (re)identify an individual either directly or in combination with other extant data.

**Artifact:** an artificial object (e.g. computer program, document, method, model, practice, technique, template).

**Chain-of-evidence:** (in qualitative research) a mapping of raw data (e.g. quotations) to theoretical concepts (e.g. themes, categories), typically with one or more intermediate steps (e.g. codes, labels, subcategories), sometimes presented as a table.

**Credibility:** the extent to which conclusions are supported by rich, multivocal evidence.

**Construct validity:** Do the measures support the research objective? The questionnaire items (questions) and related response scales should accurately represent the research aims.

**Deception:** A situation where some aspects of a study are intentionally concealed from participants to permit aspects of the research that would not be possible if full information was given.

**De-identification:** the process of removing identifying elements of data; often used synonymously with both Anonymization and Pseudonymization even though these processes have important differences. It is usually better to specify whether data is *anonymous* or *pseudonymous*.

**External validity:** Can the conclusions be generalized to the target population? The characteristics and size of the sample should represent the extended population.

**Gatekeeper:** A person or organization who controls access to individuals (e.g. employees) or data being researched. Gatekeepers may influence individual's participation decisions (e.g. employers may coerce employees to participate), affect research outcomes (e.g. by controlling whose voices are heard) or be affected by research outcomes (e.g. lose credibility with employees).

**Generalizability:** See **External validity**.

**Informed consent:** A freely-made decision by an individual to participate in research in the light of full information about the purposes, benefits, and risks to them of participating in the research, and the options available to them to withdraw themselves or their data during or after the study. The freedom to consent/withdraw may be affected by pre-existing relationships with the researchers (e.g. family/friendship) or gatekeepers.

**Internal validity:** Are the relationships between the investigated factors examined? In survey research, it is difficult to control the conditions in which the factors are studied and to account for potential confounding factors. Low internal validity is expected.

**Multivocal:** The property of being based on—and recognizing differences between—people with different opinions and backgrounds (including gender, culture, education, and class).

**Pseudonymization:** The process of removing directly-identifying elements of data and creating a separate explicit (e.g. allocating a random number to a participant's record in place of their name) or implicit (e.g. relying on timestamps to resolve logs to sign-on information) map between the identifiable aspects of the data and the remainder. *Pseudo*nymizing differs from *anon*ymizing in that individuals can be re-identified using the map.

**Objectivity:** Are the results free from the bias of the researchers? This can be achieved through standardization of the procedures for data collection, analysis, and interpretation.

**Recoverability:** A study is recoverable when readers can understand how the work was done and why it was done that way. All research should be recoverable.

**Reflexivity**: the extent to which authors reflect on their potential biases and interactions with the team, organization or community, especially possible negative impacts on some participants or stakeholders.

**Re-identification:** the process of connecting ostensibly anonymous or pseudonymous data back to an individual, typically by combining field values from a single dataset, or by combining elements from multiple datasets.

**Reliability:** The degree to which a measure can be applied consistently across time (test-retest), or when applied by different people (inter-rater).

**Replicability:** A study is replicable, when the data collection and analysis (on the new data) can be repeated by an independent researcher. Positivist research should be replicable; interpretivists and postmodernists reject the notion that social science is replicable. Qualitative research is typically not replicable.

**Reproducibility:** A study is reproducible when an independent researcher can precisely recreate the results using the original study's data and source code. Interpretivists and postmodernists reject the notion that social science is reproducible, and qualitative research is typically not reproducible. Much positivist research is not reproducible because it is impractical or unethical to publish the dataset.

**Resonance:** the extent to which a study's conclusions make sense to (i.e. resonate with) participants

**Rigor:** the extent to which theory, data collection, and data analysis are sufficient, appropriate and not oversimplified.

**Site:** the conceptual space within which a study's data collection occurs. In qualitative and especially case study research, the concept of site is not limited to physical location, but rather defines the boundaries within which the research takes place.

**Theoretical sampling:** choosing which data to collect based on the emerging theory, concepts or categories; typically used in qualitative research, especially Grounded Theory.

**Transferability:** the extent to which a study's results could plausibly apply to other sites, people or circumstances.

**Usefulness:** the extent to which a study provides actionable recommendations to researchers, practitioners **OR** educators.

# List of Contributors

Editor

**Paul Ralph**, *PhD (British Columbia), B.Sc. / B.Comm (Memorial),* is an award-winning scientist, author, consultant and Professor of Software Engineering at Dalhousie University in Halifax, Canada. Paul co-chaired the ACM SIGSOFT Paper and Peer Review Quality Initiative and has written extensively on research methodology for software engineering.

Content Contributors

**Nauman bin Ali**, Blekinge Institute of Technology, Sweden [Quantitative Simulations]

**Sebastian Baltes**, University of Adelaide, Australia [Sampling]

**Domenico Bianculli**, University of Luxemburg, Luxemburg [Engineering]

**Jessica Diaz**, Technical University of Madrid, Spain [IRR/IRA]

**Yvonne Dittrich**, IT University of Copenhagen, Denmark [Action Research]

**Neil Ernst**, University of Victoria, Canada [Pre-registration]

**Michael Felderer**, University of Innsbruck, Austria [Action Research, Case Study]

**Robert Feldt**, Chalmers University of Technology, Sweden [General]

**Antonio Filieri**, Imperial College London, UK [Engineering]

**Breno Bernard Nicolau de França**, University of Campinas, Brazil [Quantitative Simulations]

**Carlo Alberto Furia**, USI Lugano, Switzerland [Engineering]

**Greg Gay**, Chalmers University / University of Gothenburg, Sweden [Optimization Studies]

**Nicolas Gold,** University College London, UK [Ethics]

**Daniel Graziotin**, University of Stuttgart, Germany [Open Science]

**Pinjia He**, ETH Zurich, Switzerland [General]

**Rashina Hoda**, Monash University, Australia [Grounded Theory]

**Natalia Juristo**, Technical University of Madrid, Spain [Controlled Experiment]

**Barbara Kitchenham**, Keele University, UK [Systematic Reviews]

**Valentina Lenarduzzi**, LUT University, Finland [Longitudinal Studies]

**Jorge Martínez**, Universidad Politécnica de Madrid, Spain [IRR/IRA]

**Jorge Melegati,** Free University of Bozen-Bolzano, Italy [Case Survey]

**Daniel Mendez**, Blekinge Institute of Technology, Sweden, and fortiss GmbH, Germany [Questionnaire]

**Tim Menzies**, NC State University, USA [Exploratory Data Science; Optimization Studies]

**Jefferson Molleri**, Simula Metropolitan Centre for Digital Engineering, Norway [Questionnaire]

**Dietmar Pfahl**, University of Tartu, Estonia [Quantitative Simulations]

**Romain Robbes**, Free University of Bozen-Bolzano, Italy [Data Science]

**Daniel Russo**, Aalborg University, Denmark [Longitudinal Studies]

**Nyyti Saarimäki**, Tampere University, Finland [Longitudinal Studies]

**Federica Sarro**, University College London, UK [Optimization Studies]

**Davide Taibi**, Tampere University, Finland [Longitudinal Studies]

**Janet Siegmund**, Chemnitz University of Technology, Germany [Pre-registration]

**Diomidis Spinellis**, Athens University of Economics and Business, Greece [Systematic Reviews]

**Miroslaw Staron**, University of Gothenburg, Sweden [Action Research]

**Klaas Stol**, University College Cork, Ireland [Case Study]

**Margaret-Anne (Peggy) Storey**, University of Victoria, Canada [Multimethodology and Mixed Methods]

**Davide Taibi**, Tampere University, Finland [Longitudinal Studies]

**Damian Tamburri**, Eindhoven University of Technology, Netherlands [Case Study; Exploratory Data Science]

**Marco Torchiano**, Polytechnic University of Turin, Italy [Information Visualization, Questionnaire]

**Christoph Treude**, University of Adelaide, Australia [Grounded Theory]

**Burak Turhan**, Monash University, Australia [Controlled Experiment]

**Xiaofeng Wang**, Free University of Bozen-Bolzano, Italy [Case Survey]

**Sira Vegas**, Technical University of Madrid, Spain [Controlled Experiment]

## Technical Contributors

**Arham Arshad**, Dalhousie University, Canada

**Taher Ghaleb**, Queen's University, Canada